\documentclass[10pt]{bmc_article}

\usepackage{cite} 
\usepackage{url}  
\usepackage{ifthen}  
\usepackage{multicol}   
\usepackage[utf8]{inputenc} 
\usepackage{graphicx}
\usepackage[table]{xcolor}
\usepackage{fancyref}
\usepackage{rotating}
\usepackage{draftwatermark}
\SetWatermarkFontSize{64pt}
\SetWatermarkText{PREPRINT ONLY}
\SetWatermarkLightness{0.75}
\SetWatermarkScale{2.0}

\newcommand{\CEL}{\textit{C. elegans}}
\newcommand{\PPA}{\textit{P. pacificus}}
\newcommand{\RCU}{\textit{R. culicivorax}}
\newcommand{\TSP}{\textit{T. spiralis}}

\newcommand{\TCA}{\textit{T. castaneum}}

\newboolean{publ}

\newenvironment{bmcformat}{\baselineskip20pt\sloppy\setboolean{publ}{false}}{\baselineskip20pt\sloppy}

\begin{document}
\begin{bmcformat}

\begin{center}
\Huge{Major changes in the core developmental pathways of nematodes: \textit{Romanomermis culicivorax} reveals the derived status of the \textit{Caenorhabditis elegans} model}
\end{center}

\bigskip
\bigskip
\noindent Philipp H. Schiffer{$^1$$^\dag$}, Michael Kroiher{$^1$}, Christopher Kraus{$^1$}, Georgios D. Koutsovoulos$^2$,Sujai Kumar$^2$, Julia I. R. Camps$^1$, Ndifon A. Nsah$^1$, Dominik Stappert$^3$, Krystalynne Morris$^4$, Peter Heger$^1$, Janine Altm\"uller$^5$, Peter Frommolt$^5$, Peter N\"urnberg$^5$, W. Kelley Thomas$^4$, Mark L. Blaxter$^2$ and Einhard Schierenberg$^1$
\newline

$^\dag$ corresponding author: PHS - ORCiD:0000-0001-6776-0934 - p.schiffer@uni-koeln.de

\begin{enumerate}
\item Zoologisches Institut, Universit\"at zu K\"{o}ln, Cologne, NRW, Germany
\item Institute of Evolutionary Biology, School of Biological Sciences, The University of Edinburgh, Edinburgh, Scotland, UK
\item Institute f\"ur Entwicklungsbiologie, Universit\"at zu K\"{o}ln, Cologne, NRW, Germany
\item Hubbard Center for Genome Studies, University of New Hampshire, Durham, NH, USA
\item Cologne Center for Genomics, Universit\"at zu K\"{o}ln, Cologne, NRW, Germany
\end{enumerate}%

\newpage



\begin{abstract}

\subsection*{Background}
Despite its status as a model organism, the development of \textit{Caenorhabditis elegans} is not necessarily archetypical for nematodes. The phylum Nematoda is divided into the Chromadorea  (indcludes \CEL) and the Enoplea. Compared to \CEL, enoplean nematodes have very different patterns of cell division and determination. Embryogenesis of the enoplean \textit{Romanomermis culicivorax} has been studied in great detail, but the genetic circuitry underpinning development in this species is unknown.
\subsection*{Results}
We created a draft genome of \RCU\ and compared its developmental gene content with that of two nematodes, \CEL\ and \textit{Trichinella spiralis} (another enoplean), and a representative arthropod \textit{Tribolium castaneum}. This genome evidence shows that \RCU\ retains components of the conserved metazoan developmental toolkit lost in \CEL. \TSP\ has independently lost even more of the toolkit than has \CEL. However, the \CEL\ toolkit is not simply depauperate, as many genes essential for embryogenesis in \CEL\ are unique to this lineage, or have only extremely divergent homologues in \RCU\ and \TSP. These data imply fundamental differences in the genetic programmes for early cell specification, inductive interactions, vulva formation and sex determination.
\subsection*{Conclusions}
Thus nematodes, despite their apparent phylum-wide morphological conservatism, have evolved major differences in the molecular logic of their development. \RCU\ serves as a tractable, contrasting model to \CEL\ for understanding how divergent genomic and thus regulatory backgrounds can generate a conserved phenotype. The availability of the draft genome will promote use of \RCU\ as a research model.

\end{abstract}

\newpage

\ifthenelse{\boolean{publ}}{\begin{multicols}{2}}{}


\section*{Background}
Species in the phylum Nematoda have a generally conserved body plan. The classic nematode form is dictated by the presence of a hydroskeleton, where longditudinal muscles act against an inextensible extracellular cuticle. What is more surprising is the conservation of organ systems between nematode species, with, for example, the nervous system and the somatic gonad and vulva having very similar external and cellular morphologies. It might be thought that these similar morphologies and cellular structures arise from highly stereotypical developmental programmes, but observational data are emerging that challenge this "all nematodes are equal" view.
The embryonic development of the nematode \textit{Caenorhabditis elegans} (Rhabditina, Rhabditda, Chromadorea; see De Ley and Blaxter \cite{Blaxter:1998wj}) has become a paradigmatic model for studying developmental processes in animals, including earliest soma-germline separation, fate specification through cell-cell interactions, and differentiation.The particular mode of development of \CEL\ is distinct within the major metazoan model organisms, but much of the regulatory logic of its development is comparable to that in other phyla. One key aspect in which \CEL\ differs from vertebrate and arthropod models is that \CEL\ has a strictly determined developmental programme \cite{Maduro:2010hd}, with a largely invariant cell lineage giving rise to predictable sets of differentiated cells \cite{Sulston:1983vv}. Inductive cell-cell interactions are, nevertheless, essential for its correct development \cite{Maduro:2010hd}.
The first description of the early embryogenic cell lineage of a nematode, that of \textit{Ascaris} (Spirurina) in the 1880's \cite{Boveri:1899us, Muller:1903tj}, conforms to the \CEL\ model. 

Early development across all three suborders of the Rhabditida (i.e. Rhabditina, Tylenchina and Spirurina sensu De Ley and Blaxter \cite{Blaxter:1998wj}) is very similar \cite{Vangestel:2008uj, Skiba:1992ut}. In general only relatively minor variations on the division pattern observed in \CEL, including heterochrony in the timing of particular cell divisions, and restrictions in cell-cell interaction due to different placement of embryonic blastomeres within the eggshell following altered orientations of cell division spindles have been described in these nematodes \cite{Lahl:2009cw, Brauchle:2009fc}. From this large body of work it might be assumed that all nematodes follow a \textit{C. elegans}-like pattern of development. Deviations from the \textit{C. elegans} pattern observed in these rhabditid nematodes indicate that the determined mode of development is subject to evolutionary change, and have assisted in revealing the underpinning regulatory logic of the system. Indeed, a greater role for regulative interactions in early development has been characterised in some rhabditids, such as \textit{Acrobeloides nanus} (Tylenchina) \cite{Wiegner:1998ut, Wiegner:1999jw}.\\
Regulative development is common in Metazoa, and is also observed in other ecdysozoan taxa (e.g. within the Arthropoda). The determined mode found in \CEL\ is thus likely to be derived. 
Molecular and morphological systematics of the phylum Nematoda identify two classes: Chromadorea (including Rhabditida), and Enoplea (subdivided into Dorylaimia and Enoplia) \cite{Blaxter:1998wj, Meldal:2007te} (Figure 1). In several Enoplea, early embryos do not display polarised early divisions, and observational and experimental evidence argues against a strongly determined mode of development \cite{Voronov:1998vy, Schulze:2011gh}. Strongly determinative development may thus be derived even within Nematoda \cite{Schulze:2012du}. This implies that the underpinning developmental system in Nematoda has changed, while maintaining a very similar organismal output. This phenomenon, termed 'developmental system drift' \cite{True:2001tp}, allows independent selection on the mechanism and the final form produced by it. To explore mechanistic aspects of development of enoplean and other non-rhabditid nematodes requires tractable experimental systems with a wealth of underpinning methodological tools and extensive genetic data. While \CEL\ and its embryos are relatively easily manipulated and observed, and the \CEL\ genome has been fully sequenced \cite{CelegansSequencingConsortium:1998wf}, embryos from taxa in Enoplia and Dorylaimia are much harder to culture and manipulate. Few viable laboratory cultures exist and obtaining large numbers of embryos from wild material is difficult. Functional molecular analyses of species in most nematodes, and Enoplia and Dorylaimia in particular, is further hindered by the lack of genetic tools such as mutant analysis or gene-knockdown via RNAi.\\
While realisation of extensive programmes of comparative experimental embryology across the phylum Nematoda remains a distant research goal, we have taken a parallel genome-based approach. Using the background knowledge of pathways and modules used in other taxa, the underpinning logic of a species' developmental system can be inferred from its genome, and the developmental toolkits of different species can be compared. These comparisons can pinpoint changes in developmental logic between taxa by identifying genes unique to one species or group, and gene losses during evolution, that must result in changed pathway functioning. Efficient generation of genomic resources for non-model species, and the inference of developmental regulatory pathways from the encoded gene sets, is now possible. The majority of the 11 genome sequences determined to date for Nematoda has been from Rhabditida (e.g. \CEL\ and congeners) \cite{Stein:2003ks, Mortazavi:2010ki, Dieterich:2008bi, Jex:2011ew, Ghedin:2007db, Godel:2012iz, Abad:2008kv, Kikuchi:2011fz}. A single member of Enoplea, the mammalian parasite \textit{Trichinella spiralis} (Dorylaimia; order Trichocephalida) has been sequenced \cite{Mitreva:2011ik}. \TSP\ is ovoviviparous, and proper development requires the intrauterine environment. \TSP\ blastomeres are extremely transparent  \cite{Hope:2002wn} such that individual nuclei are hard to identify (E.S., unpublished observations). Hence this species is of very limited value for image analysis and experimental investigations correlating cellular aspects and the underpinning molecular logic of early development. The genomes of many additional nematode species are being sequenced and annotated \cite{Kumar:2011cr,Kumar:2012wg}, but even in this wider sampling of the phylum, Enoplia and Dorylaimia are neglected.\\

\textit{Romanomermis culicivorax} (order Mermithida within Dorylaimia), has been established in culture for decades. \RCU\ infects and kills the larvae of many different mosquito species \cite{petersen:1985wx}, and is the subject of research programmes investigating its potential as a biocontrol agent of malaria and other disease vectors \cite{Petersen:1978wb, petersen:1985wx}. \RCU\ and \TSP\ differ fundamentally  in many life-cycle and phenotypic characters. Free living \RCU\ juveniles actively seek and invade mosquito larvae in the water \cite{Shamseldean:1989tx}, while \TSP\ is transmitted as an arrested, first stage larva encysted in muscle tissue \cite{Mitreva:2006wb}. \textit{R. culicivorax} embryos are easily studied under laboratory conditions, and a single female can produce more than a thousand eggs in culture. They display a developmental pattern that differs markedly from the \CEL\ model. As in other Dorylaimia and Enoplea \cite{Schulze:2008bc, Schulze:2011gh} the first division is equal, and not asymmetric as in \CEL. \RCU\ also shows an inversion of dorso-ventral axis polarity compared to \CEL. A predominantly monoclonal fate distribution in \RCU\ somatic founder cells indicates fewer modifying inductions between blastomeres \cite{Schulze:2008bc, Schulze:2009kb}. Generation of the hypodermis involves repetitive cell elements extending from posterior to anterior over the remainder of the embryo, a system very different from that of \CEL\ \cite{Schulze:2009kb}. In the context of this distinct developmental mode in \RCU, we decided to catalogue its developmental toolkit by sequencing the genome, and here present a draft assembly and annotation. We contrast the toolkits identified in \RCU\ and \TSP\ with that of \CEL, and of other metazoa, notably the arthropod \textit{Tribolium castaneum}. We conclude that major changes in the regulatory logic of development have occurred during the evolution of nematodes, possibly as a consequence of developmental system drift, and that the model species \CEL\ represents an extreme derivation from a shared metazoan ground system.

\section*{Results and Discussion}
\subsection*{\textit{Romanomermis culicivorax} has a large and repetitive genome}
A draft genome assembly for the mermithid nematode \textit{R. culicivorax} was generated from 26.9 gigabases (Gb) of filtered raw data (from a total of 41 Gb sequenced; Table 1). The assembly has a contig span of 267 million base pairs (Mb) and a scaffold span of 323 Mb. The 52 Mb of spanned gaps are likely inflated estimates derived from our use of the SSPACE scaffolder. We do not currently have a validated independent estimate of genome size for \RCU, but preliminary measurements with Feulgen densitometry suggest a size greater than 320 Mb (Elizabeth Mart\'inez Salazar pers. comm.). The \RCU\ genome is thus likely to be three fold bigger than that of \CEL, and five fold that of \TSP\ (Table 2). 
The assembly is currently in 62,537 scaffolds and contigs larger than 500 bp, with an N50 of 17.6 kb. The N50 for scaffolds larger than 10 kb is 29.9 kb, and the largest scaffold is over 200 kb. The GC content is 36.3\%, comparable to 38\% of \CEL\ and 34\% in \TSP. We identified 47\% of the \RCU\ genome as repetitive. To validate this estimate we repeated our repeat-finding approach against previously published genomes and achieved good accordance with published data (Table 2). The non-repetitive content of the \RCU\ genome is thus approximately twice that of \CEL\ and three times that of \TSP. \TSP\ thus stands out as having the least complex nematode genome sequenced thus far, and the contrast with \RCU\ shows that small genomes are not a characteristic of Dorylaimia.\\
The RNA-Seq data were assembled into 29,095 isotigs in 22,418 isogroups spanning 23 Mb, and thus are likely to be a reasonable estimate of the  \RCU\ transcriptome. Using BLAT \cite{Kent:2002tv}, 21,204 of the isotigs were found to be present (with matches covering $>$80\% of the isotig) in single contigs or scaffolds of the genome assembly, suggesting reasonable biological completeness and contiguity. We also used the CEGMA approach to assess quality of the genome assembly, and found high representation (89.92\% partial, 75.40\% complete) and low proportion of duplicates (1.05 fold), suggesting a high quality assembly with limited retained haploid assembly duplicates (Table 1). Automated gene prediction from the assembly with iterative rounds of the MAKER pipeline, using the RNA-Seq data as evidence both directly and through GenomeThreader-derived mapping, yielded a total of nearly 50,000 gene models. These were reduced to 48,171 gene models by merging those with identities $>$99\% using Cd-hit. This gene count would be surprisingly high for a nematode: \CEL\ has $\sim$22,000 genes, \TSP\ has $\sim$16,000, and \textit{Pristionchus pacificus} has $\sim$27,000. The excess of \RCU\ gene models may result from poorly assembled contigs, from assembly fragmentation, and "over-enthusiastic" prediction from gene modelers within the MAKER pipeline. Within the 48,171 predictions, 12,026 were derived from the Augustus modeler and 36,145 from SNAP.  Because Augustus predictions conservatively require some external evidence (transcript mapping and/or sequence similarity to other known proteins), we regarded these as the most reliable and biologically complete.
Exons of the Augustus-predicted genes in \RCU\ had a median length of 161 bp, slightly larger than those in \CEL\ (137bp) and \TSP\ (128bp). Introns of the \RCU\ Augustus models, with a median of 405 bp, were much larger than those in \CEL\ (69 bp) or \TSP\ (283bp). The small introns observed in \CEL\ and other rhabditid nematodes (Table 2) are thus likely to be a derived feature.\\
We annotated 1,443 tRNAs in the \RCU\ genome using INFERNAL \cite{Nawrocki:2009kr} and tRNAscan-SE \cite{Lowe:1997uc}, of which 382 were pseudogenes (see Table S5 for details). In comparison, \TSP\ has 134 tRNAs of which 7 are pseudogenes, while \CEL\ has 606 tRNAs with 36 pseudogenes \cite{DataforPhDThesis:2012fd}. Threonine (Thr) tRNAs were particularly overrepresented (676 copies), a finding echoed in the genomes of \textit{Meloidogyne incognita} and \textit{Meloidogyne floridensis} (tylenchine nematodes, see Figure 1) \cite{Abad:2008kv} and in \PPA\ \cite{Dieterich:2008bi}. \PPA\ also has an overrepresentation of Arginine tRNAs \cite{DataforPhDThesis:2012fd}.\\
We have made the annotated \RCU\ genome, with functional categorisations of predicted genes and proteins and annotation features, available in a dedicated genome browser at \url{http://romanomermis.nematod.es.}

\subsection*{The \textit{R.culicivorax}  proteome retains conserved metazoan components lost in \textit{T. spiralis} and \textit{C.elegans}}
The phylogenetic placement of \RCU\ compared to \CEL\ makes its genome ideal for exploring the likely genetic complexity of the ancestral nematode. With \TSP, it can be used to reveal the idiosyncracies of the several genomes available for Rhabditida. To polarise this comparison, we used data from the genome of the arthropod \textit{T. castaneum}. The \TCA\ genome is of high quality \cite{Richards:2008fa} and the pattern of development of this beetle is less derived than that of the major arthropod model \textit{Drosophila melanogaster} \cite{Schroder:2008kk}. We used the orthoMCL pipeline to generate a set of gene clusters for the four species \RCU, \TSP, \CEL\ and \TCA. The large sequence divergence between the four species may have obscured orthology relationships, making inference of true functional orthology problematic \cite{Jensen:2001uu, Koonin:2005tt, MorenoHagelsieb:2008ck}, but the parameters used (a BLAST E-value of 1e$^-$$^5$, and orthoMCL inflation parameter of 1.5) can be regarded as relaxed (i.e. most inclusive) compared to other studies \cite{MorenoHagelsieb:2008ck, Shaye:2011um, Tautz:2011dl}. As the \RCU\ genome assembly may not be complete, we based inference of absence on shared loss in both \RCU\ and \TSP. Thus, we believe that our analyses were at a minimum able to identify homologues where present, and thus we could robustly infer absence. While the orthoMCL pipeline is regarded as very robust in accurately clustering unknown proteins \cite{Chen:2007ft} inferences of functional or biological orthology are complex. Inferences of absence were explored in detail (Supplementary file 5).\\

We identified 3274 clusters that contained protein representatives from all three nematode genomes, and 2833 of these also contained at least one \TCA\ representative (Figure 2). These 2833 clusters represent a conserved metazoan and eukaryotic core proteome. There were many clusters that contained proteins from only one species of nematode, representing lineage specific expansions of novel protein families. \TSP\ had the lowest number of these (975), while \CEL\ and \RCU\ each had over two thousand. Interestingly, of the 2747 \RCU-limited clusters, 324 (11.8\%) had apparent orthologues in \TCA. Such clusters are candidates for retention of phylogenetically ancient genes by one nematode species and loss in the other two.\\ 
\TSP\ appeared to have lost more phylogenetically ancient genes than had either \RCU\ or \CEL. \TSP\ and \CEL\ shared only 412 clusters exclusive of \RCU\ members, while \RCU\ and \CEL\ shared 1298 clusters exclusive of \TSP. Despite their phylogenetic affinity, \RCU\ and \TSP\ only shared 600 clusters exclusive of \CEL. \CEL\ and \RCU\ shared very similar numbers of clusters with \TCA\ (2833 contain all species in the comparison; 853 contain only \CEL, \RCU\ and \TCA, 569 \CEL\ and \TCA, and 568 \RCU\ and \TCA) (Figure 2).\\

The clusters containing only \textit{R. culicivorax} and \textit{T. spiralis} might identify functions important to these dorylaim nematodes. In the 461 \textit{T. spiralis} and 806 \textit{R. culicivorax} proteins in these clusters, a total of 65 GO terms were found to be overrepresented (p$<$0.05 by Fisher's exact test) compared with the GO annotation set derived from the complete \textit{C. elegans} proteome, and 33 were overrepresented when compared to annotation of the \textit{T. castaneum} genome. There were 26 GO terms overrepresented in both comparisons. Clusters with \textit{R. culicivorax}, \textit{T. spiralis} and \textit{T. castaneum} members (but lacking \textit{C. elegans} members) contained 332 \textit{R. culicivorax} and 573 \textit{T. spiralis} and 445 \textit{T. castaneum} proteins, and we identified 40 GO terms overrepresented compared to the GO annotated \textit{C. elegans} proteome (see Supplementary file 2).
From this we suggest that \textit{T. spiralis} may not have a typical dorylaim genome. The \textit{T. spiralis} genome is reduced in content compared to other nematodes: it is smaller, has fewer genes overall, and has fewer phylogenetically ancient genes. This is congruent with the previously reported loss of proteins with metabolic function in \textit{T. spiralis} \cite{Mitreva:2011ik}. The evolutionary reasons behind this reduction remain obscure, but could include loss of genetic capacity following acquisition of a unique lifestyle that lacks a freeliving stage or genomic streamlining to permit rapid reproduction and growth. Many parasitic and endosymbiotic prokartyotes and eukaryotes have reduced genome sizes \cite{Keeling:2010dv}.\\

\subsection*{The genetic background of development in \RCU\ and \TSP\ differs markedly from that of \CEL}
In a recent multi-species developmental timecourse expression analysis within the genus \textit{Caenorhabditis}, conserved sets of genes were found to be over-expressed in discrete portions of the developmental timeline from zygote to hatching larva \cite{Levin:2012vd}. In particular, this study suggests a conserved period in development where a very restricted set of genes is expressed in all species, perhaps corresponding to a 'bauplan' stage in nematode development as has been proposed for Metazoa in general. To explore whether this model can be extended across Nematoda, we identified \RCU\ and \TSP\ homologues of the 1725 developmentally regulated \CEL\ genes extracted from this analysis \cite{Levin:2012vd}. Nearly half (845) of these genes were not grouped in clusters with Dorylaimia proteins using orthoMCL. We were unable to identify any sequence homologs for 450 of the proteins in \textit{R. culicivorax} using BLAST+.\\
The remaining 395 proteins had BLAST+ hits to \textit{R. culicivorax} proteins, but were so divergent that orthoMCL did not cluster them as orthologs with Dorylaimia proteins. Among these 395 with marginal matches, we found that 18 belonged to the \textit{C. elegans} nuclear hormone receptor subfamilies, 5 were innexin type gap-junction protein, 6 were TWiK potassium channel proteins and 5 were acetylcholine receptor proteins. These protein families are particularly diverse and expanded in \CEL\ \cite{PHELAN:2005hm,Jones:2003kl, Antebi:2006bh, Altun:2009fv} and we suggest that the genes "missing" from \RCU\ but having low-scoring BLAST+ matches represent rapidly evolved, divergent duplications within the lineage leading to \CEL. OrthoMCL is likely to be correct in not clustering most of these proteins. 
The proportion of \textit{Caenorhabditis}-restricted genes across the developmental timecourse examined by Levin et al. \cite{Levin:2012vd} varied from 36.4\% to 59.9\% (Figure 3 and Supplementary file 4). A surprisingly high proportion of the developmental genes acting during specific embryonic stage transitions appear to be unique to the genus \textit{Caenorhabditis} or at least so divergent that functional orthology, including interaction with conserved partners, is doubtful. A striking difference between \RCU\ and \TSP\ was apparent, with 238 of the developmentally differentially expressed \CEL\ genes having a \RCU\ homologue but not a \TSP\ homologue, while only 88 had a \TSP\ homologue but not an \RCU\ one. Given the conservatism of body plan evolution in nematodes, these dramatic genetic differences suggest extensive, largely phenotypically "silent" changes in the genetic programmes orchestrating nematode development. We used computational comparisons of selected key molecular processes and pathways to tease out the differences between the model \CEL\ and the two dorylaim species, \TSP\ and \RCU.\\

\subsection*{Core developmental pathways differ between nematodes}
There are important differences in the cellular biology of development between \RCU\ and \CEL\ \cite{Schulze:2008bc, Schulze:2009kb}, and we used the genomic data to follow up on some of the more striking contrasts between the dorylaim and the rhabditid patterns of development: primary axis polarity, segregation of maternal message within the early embryo, hypodermis formation, the vulval specification pathways, epigenetic pathways (especially DNA methylation), sex determination and light sensing.\\ 
In the \CEL\ 2-cell stage mitotic spindles rotate 90\% in the posterior germline cell, and the subsequent cell divisions are orthogonal \cite{Bowerman:2000gb,Severson:2003ek, Gonczy:2005ff}. This rotation is not observed in \RCU\ and division is longditudinal \cite{Schulze:2008bc}.  
In \CEL\ and many other animals \textit{par} genes are essential for cell polarisation \cite{Goldstein:2007eo} and polarised distribution of PAR proteins results in the restriction of mitotic spindle rotation to one cell. \CEL\ mutants lacking \textit{par-2} and \textit{par-3} genes resemble the \RCU\ phenotype, showing longitudinal spindle orientation \cite{Cheng:1995vl}. The \textit{par-2} gene was missing from both \RCU\ and \TSP\ (Figure 3; Table 3). Additionally, no orthologues for the \textit{par-2}-interacting genes \textit{let-99}, \textit{gpr-1} or \textit{gpr-2}, required for proper embryonic spindle orientation in \CEL\ \cite{Wu:2007jc}, were identified in the dorylaims using orthoMCL clustering or sensitive BLAST searches. 
We identified a candidate \textit{par-3} in \RCU, but this was so divergent from \CEL, \TCA\ and \TSP\ \textit{par-3} that these putative orthologues were not clustered in our analysis. The \textit{D. melanogaster} \textit{par-3} ortholog \textit{bazooka} functions in anterior-posterior axis formation, but as in \RCU\ and \TSP\ \textit{par-2} is absent from the fly \cite{Doerflinger:2010dj}. Thus, we hypothesise that the PAR-3 - PAR-2 system for regulating spindle positioning evolved within in the lineage leading to the genus \textit{Caenorhabditis}. The divergent \textit{par-3}-like gene in dorylaims may be involved in axis formation, but perhaps interacts with different partner proteins.\\
Once polarity has been established in the early \CEL\ embryo, many maternal messages are differentially segregated into anterior or posterior blastomeres \cite{Goldstein:1998wx, Gonczy:2005ff}. MEX-3 is an RNA-binding protein translated from maternally-provisioned mRNAs found predominantly in early anterior blastomeres \cite{Draper:1996tq,Huang:2002wv}. We identified a highly divergent MEX-3 homologue in \RCU, but found no orthologue in \TSP.

To demonstrate the utility of the \textit{R. culcivorax} system, and the power of the genome-to-development model, we assayed its expression in embryos using \textit{in situ} hybridisation. We selected the \textit{mex-3} gene for these studies, as it
is strongly expressed and highly localised during a short time window in development in \CEL. The observed expression pattern in \RCU\ is similar to \CEL\ (Figure 5). In the fertilized \RCU\ egg \textit{mex-3} mRNA is initially equally distributed. Prior to first cleavage \textit{mex-3} mRNA is segregated to the anterior pole and thus becomes essentially restricted to the somatic S1 blastomere (for nomenclature, see \cite{Schulze:2011gh}). With the division of S1 it is localised to both daughter cells. After the 4-cell stage the signal disappears gradually. Despite the presence, and apparent conservation of expression pattern, of \textit{mex-3}, we were unable to identify other components of the \CEL\ maternal mRNA regulation system, such as \textit{mex-5}, \textit{mex-6} and \textit{spn-4} in either dorylaim species. While MEX-5 and MEX-6 are important for controlled MEX-3 expression in \CEL\ \cite{Evans:2005id}, the apparent absence of SPN-4 in \RCU\ and \TSP\ is particularly intriguing. SPN-4 links embryonic polarity conferred by the \textit{par} genes and partners to cell fate specification through maternally deposited mRNAs and proteins \cite{Gomes:2001ve,Labbe:2002tc}. This suggests that the core regulatory logic of the early control of axis formation and cell fate specification must differ significantly between the dorylaim species and \CEL.\\
The hypodermis in \CEL\ is derived from specific descendants of the anterior S1 (AB) and the posterior S3 (C) founder cells \cite{Simske:2001dz}. In contrast, in \RCU\ the hypodermis is derived from S2 (EMS) descendants, which form repetitive ring structures that extend from posterior to anterior \cite{Schulze:2009kb}. Several developmental regulatory genes expressed in the hypodermis or associated with hypodermal development were present only in \CEL\ in our analysis (see Table 3 and Supplementary file 4). The GATA-like transcription factor  gene \textit{elt-3} gene is absent in the dorylaim species, but the \textit{elt-3} ortholog \textit{elt-1} is conserved in \RCU, \TSP\ and \TCA. These genes act redundantly in \CEL\ hypodermis formation \cite{Gilleard:2001gh}. Thus \textit{elt-3} involvement must be an innovation in the rhabditid lineage, suggesting changes of interaction complexity during nematode evolution.\\
In \CEL, vulva formation is highly dependent on the inital cell-cell interactions of the anchor cell with the neighboring vulva precursor cells (VPCs). Induction of the VPCs activates a complex gene regulatory network which drives divisions and differentiations of the VPCs to form a functional vulva. The evolutionary lability of this system has been explored in rhabditid nematodes, revealing the changing relative importances of a series of cell-cell interactions, short and long range inductions, and lineage-autonomous specifications \cite{Sommer:2001up, Kiontke:2007jj}. The signal transduction pathways involved include a RTK/RAS/MAPK cascade, activated by EGF- and wnt-signaling \cite{Sternberg:2005fx}. Among the downstream targets in \CEL\ are for example \textit{lin-1} and the $\beta$-catenin \textit{bar-}1, which in turn regulates the HOX-5 ortholog \textit{lin-39} \cite{Salser:1993uk, Eisenmann:2000wh, Shemer:2002hy}. Our analysis shows that \textit{lin-1} and \textit{bar-}1, as well as other important regulators of vulva development, are absent from the genomes of \RCU\ and/or \TSP\ (Table 3 and Supplementary File 4). We identified a \textit{R. culicivorax} gene with a low-quality match to BAR-1 (24.2\% sequence identity). This protein is not clustered with other dorylaim proteins, and appears to be either a duplication of the $\beta$-catenin ortholog HMP-2 or another armadillo repeat-containing protein and not orthologous to \textit{bar-1} (see Supplementary file 5). These shared patterns of absence again indicate that the same morphological structures can be generated with very different genetic underpinnings. While it is possible that vulva formation in the dorylaims is regulated without the \textit{bar-1} - \textit{lin-39} interactions, as observed in \PPA\ \cite{Tian:2008jh}, it may be that HOX genes function differently in the dorylaims: rather than acting in a lineage-dependent manner (as in \textit{C. elegans} \cite{Streit:2002vo, Aboobaker:2003wx}) they may act in a positional regulatory manner, as in other animals \cite{Aboobaker:2003uu, Lemons:2006ct}.\\
Epigenetic regulation is key to developmental processes in many animals, but its roles in \CEL\ are more muted. While \CEL\ has a reduced ability to methylate DNA \cite{Bird:2002hs}, orthologue clusters restricted to \RCU\ and \TSP\ (excluding \CEL) were enriched for four methylation-associated GO terms. We also found significant enrichment (p$<$0.05) for GO terms describing chromatin and DNA methylation functions in the set of \RCU\ proteins that lacked homologues in \CEL\ (see Supplementary file 2). Important roles for methylation and changes in methylation patterns in the development of \TSP\ have been inferred from transcriptional profiling \cite{Gao:2012fm}. In addition, methylation is important for the silencing of transposable elements \cite{Tran:2005ji, Martienssen:2001jj} and could play a crucial role in the highly repetitive \RCU\ genome. The \CEL\ genome was also found to be depleted for chromatin re-modeling genes of the Polycomb and Trithorax groups \cite{Chamberlin:2000ta}. It is intriguing that we found orthologs of \TCA\ \textit{pleiohomeotic} in \RCU\ and \TSP, and orthologs of \TCA\ \textit{trithorax} and \textit{Sex comb on midleg} (\textit{Scm}) in \RCU. This suggests that dorylaim chromatin restructuring mechanisms may be much more arthropod-like than are those of \CEL. The presence of an intact methylation machinery and conserved chromatin re-modelling factors opens the prospects for a role for epigenetic modification in developmental regulation in dorylaim nematodes.

\subsection*{Sex determination machinery}
The mechanism of sex determination differs considerably among animals and it has been claimed to be one of the developmental programs most influenced by developmental system drift \cite{True:2001tp}. Sex ratios in \RCU\ are described to be environmentally determined through in-host nematode density \cite{Tingley:1986uc}, and thus might be fundamentally different from the system found and extensively analysed in \CEL \cite{Haag:2006tr}. Environmental sex determination is found in many nematode taxa, including Strongyloididae and Meloidogyninae (both Tylenchina), taxa more closely related to \CEL. \CEL\ sex determination is based on the X to autosome ratio, with males haploid for the X chromosome (XO), and females diploid (XX). This difference is read by the master switch \textit{xol-1} \cite{Powell:2005ej}, which acts through the three \textit{sdc} genes \cite{Chu:2002iy, Meyer:2005ur, Zarkower:2006kk} to regulate the systemic secretion of HER-1, a ligand for the TRA-2 receptor \cite{Kuwabara:1992th, Goodwin:2002tp, Baldi:2009hj}. TRA-2 in turn negatively regulates a complex of \textit{fem} genes, which regulates nuclear translocation of TRA-1, the final shared step in the pathway that switches between male and female systems. We did not find credible homologues (through orthoMCL and re-confirmation with BLAST) of \textit{xol-1}, \textit{sdc-1}, \textit{sdc-2}, \textit{sdc-3}, \textit{her-1} or \textit{tra-2} in either \TSP\ or \RCU\ (Table 3; Supplementary file 5), and thus these species are unlikely to use the HER-1/TRA-2 ligand-receptor system to coordinate organism-wide sexual differentiation.

\subsection*{Light sensing machinery in \RCU}
Light sensing with and without eye-like organs has been described in other mermithids \cite{Mohamed:2007wg,Robinson:1990uq}. Although \RCU\ has no structurally evident eye spots it is likely that invasion of the mosquito host on the surface of the water body \cite{Shamseldean:1989tx} and migration of emerged nematodes migrate back to the substrate to mate and deposit eggs \cite{Shamseldean:2007wl} involves phototactic behaviour. Preliminary experiments with \RCU\ give support to this view (J. Burr, pers. comm.), but the underlying physiology has not been explored. We identified several GO terms associated with photoreceptor development and light sensing (see Supplementary file 2) in \RCU\ proteins in comparison to \CEL\ and \TCA\ proteomes (in the set of \RCU\ proteins without homologues in these species). Two especially intriguing GO terms were 'phototaxis' and 'energy taxis'. Proteins associated with these GO terms had BLAST similarities to COUP transcription factors, which in the mouse have been associated with cell fate determination in the eye \cite{Tang:2010gv}.\\
In \textit{Mermis nigrescens}, a close relative of \RCU, a directional light sensing organ is found in the anterior pharynx, where a cylinder of light-shadowing cells packed with a nematode hemoglobin shades a central photoreceptor \cite{AHJBurr:2000ta, Burr:2000ue, Mohamed:2007wg}. Globin-like domains are found in diverse gene families in Nematoda \cite{Hunt:2009wf}. More \textit{R. culcivorax} proteins were annotated with the  GO term 'oxygen binding' than those of the other species analysed (Supplementary figure 1). Several of these \RCU\ proteins have BLAST matches to bonafide globins and hemoglobins, and an optical shadowing function is possible for one or more of them. Pigment granules are segregated into the hypodermis of \RCU\ (see Figure 5) and may also have a light-shadowing function \cite{Schulze:2008bc}. We also found the GO term 'cellular pigment accumulation' in the set of \RCU\ proteins that had homologues with \TSP\ and \TCA, but not with \CEL. The protein associated with this GO term was most similar to \textit{Xenopus} SHROOM2 protein, which is involved in melanosome formation and expressed in the eye of the frog \cite{Fairbank:2006ha}. We also identified a candidate opsin in \textit{R. culicivorax}. The gene is partially supported by EST data, and could generate a 313 amino acid protein with identities of 26\% to the \textit{Bos taurus} (accession NP\_776991) and \textit{Didelphis aurita} (ABC75817) long-wave-sensitive opsins.

\section*{Conclusions}

By combining the \RCU\ genome presented here together with the published \TSP\ genome, we have been able to explore the molecular diversity of of Dorylaimia, and provide robust contrasts with the intensively studied Rhabditida. Particularly surprising were the differences between \RCU\ and \TSP. The \RCU\ genome is much larger than that of \TSP. A majority of the genome was identified as repetitive, including many transposable elements. Despite the phylogenetic and lifestyle affinities between the two dorylaims compared to \CEL, the \RCU\ genome retained many more genes in common with \CEL\ than did \TSP. We suggest that \TSP\ may be an atypical representative of dorylaim nematodes, perhaps due to a highly derived life cycle.\\
Our analyses identified many genes apparently absent from the dorylaim genomes. We used very relaxed anaysis parameters, and performed close analyses of  genes identified as critical in \CEL\ development for which we could find no credible dorylaim orthologues. In these phylum-spanning comparisons, inferences of gene orthology can be obscured by levels of divergence. In addition, the gene family birth rate in the chromadorean lineage leading to \CEL\ is high \cite{Mitreva:2011ik, Kikuchi:2011fz}, and therefore \CEL\ was expected to have many genes absent from the dorylaim species. Thus, we might not have found a \RCU\ orthologue for a specific gene for three reasons: it may have arisen in the branch leading to \CEL; its sequence divergence may be too great to permit clustering with potential homologs; or it was not assembled in the draft dorylaim genomes. The analyses of \CEL\ PAR-3 and \textit{D. melanogaster} \textit{bazooka} illustrate some of these difficulties: the possible \RCU\ orthologue was highly divergent. Whether or not we have been able to identify all the orthologues of the key \CEL\ genes present in the \RCU\ and \TSP\ genomes, the absence of an identified orthologue maximally implies loss from the genome, and minimally implies significant sequence, and thus functional, divergence.\\
Between the model organisms \CEL\ and \textit{D. melanogaster} many key mechanisms governing early cell patterning are divergent \cite{Bowerman:2000gb}. Our data indicate that a major divergence also exists within Nematoda. \textit{T. spiralis} and \textit{R. culicivorax} share a lack of orthologues of genes involved in several core developmental processes in \CEL, and many of these \textit{C. elegans} genes are restricted to the Rhabditida. It is thus doubtful that these processes are regulated by same molecular interactions across the phylum. To the contrary it is likely that developmental system drift has played (and still plays) a major role in nematode evolution. The phenotypic conservatism associated with the vermiform morphology of nematodes \cite{deLey:2006jd} has fostered unjustified expectations concerning the genetic programmes that determine these morphologies.\\
To be useful as a contrasting system to the 'canonical' \CEL\ model, any nematode species must be accessible to both descriptive and manipulative investigation. Here, we have defined a reference genome for \RCU, laying bare the core machinery available for developmental regulation, and demonstrated that \textit{in situ} hybridisation approaches are feasible for this species. Along with the robust laboratory cultures long established, this makes \RCU\ an attractive and tractable alternative model for understanding the evolutionary dynamics of nematode developmental biology.
We have highlighted a few of the possible avenues a research programme could follow: early axis formation and polarisation, the specification of hypodermis, sex determination, vulva formation and the roles of epigenetic processes in developmental regulation. The advent of robust, affordable and rapid genome sequencing also opens the vista of large-scale comparative genomics of development across the phylum Nematoda \cite{Kumar:2011cr} to better understand the diversity of the phylum and also place the remarkable \CEL\ model in context of its peers. It will next be necessary to extend these analyses to a broader sampling of developmental pathway genes from a wider and fully representative sampling of nematode genomes across the full diversity of the phylum. 

\section*{Methods}
\subsection*{Sequencing and Genome Assembly}
Genomic DNA was extracted from several hundred, mixed-sex, adult \RCU\ specimens from a culture first established in Ed Platzer's laboratory in Riverside, California. Illumina paired end and mate pair sequencing with libraries of varying insert sizes, and Roche 454 single end sequencing, was performed at the Cologne Center for Genomics - CCG (\url{http://www.ccg.uni-koeln.de}). A Roche 454 dataset of transcriptome reads from cDNA synthesised from mixed developmental stages and sexes was also generated (see Table S1 for details of data generation).\\
The quality of the raw data was assessed with FastQC (v.0.9) (\url{http://www.bioinformatics.babraham.ac.uk/projects/fastqc/}). Adapter sequences and low quality data were trimmed from the Illumina paired end data with custom scripts (see \url{http://github.com/sujaikumar/assemblage}) and from the mate pair libraries with Cutadapt (v.1.0) \cite{Martin:2011va}. We constructed a preliminary genome assembly, with relaxed insert size parameters, from the paired end Illumina libraries with the de-novo-assemble option of the clcAssemblyCell (v.4.03b) \cite{Whitepaperondeno:uq}. We validated the actual insert sizes of our libraries by mapping back the reads to this preliminary assembly using clcAssemblyCell. The preliminary assembly was also used to screen out bacterial and other contaminant data \cite{Kumar:2012ez}. The transcriptome data were assembled with Roche GSAssembler (Newbler; version 2.5). For the production assembly, we explored assembly parameters using different mixes of our data, evaluating each for total span, maximal contig lengths, N50, number of contigs, representation of the transcriptome, and conserved eukaryotic gene content (using the CEGMA pipeline in version 2.1 \cite{Parra:2007df}). The most promising assembly was scaffolded with the filtered Illumina mate pair read sets using SSPACE (v.1.2) \cite{Boetzer:2011gr}. As our genomic DNA derived from a population of nematodes of unknown genetic diversity, we removed short contigs that mapped entirely within larger ones using Cd-hit (v.4.5.7) \cite{Li:2006us} at a 95\% cutoff. A final round of superscaffolding was performed, linking scaffolds that had logically consistent matches to the transcriptome data based on BLAT \cite{Kent:2002tv} hits and processed with SCUBAT (B. Elsworth, pers. comm.; \url{http://github.com/elswob/SCUBAT}). The final genome assembly was again assessed for completeness by assessing the mapping of the transcriptome contigs and with the CEGMA pipeline \cite{Parra:2007df}.

\subsection*{Genome Annotation}
RepeatMasker (v.3.3.0) \cite{RepeatMaskerOpen:ov95ByFr, Jurka:2005bl}, RepeatFinder \cite{Volfovsky:2001wa} and RepeatModeler (v.1.0.5) (\url{http://www.repeatmasker.org/RepeatModeler.html}; combining RECON (v.1.07) \cite{Bao:2002he} and RepeatScout (v.1.05) \cite{Price:2005bo}), were used to identify known and novel repetitive elements in the \RCU\ genome. We employed the MAKER pipeline\cite{Cantarel:2008jo} to find genes in the \RCU\ genome assembly. In a first pass, the SNAP gene predictor included in MAKER was trained with a CEGMA \cite{Parra:2007df} derived output of predicted highly conserved genes. As additional evidence we included the transcriptome assembly and a set of approximately 15,000 conserved nematode proteins derived from the NEMBASE4 database\cite{Parkinson:2004tv} (recalculated by J. Parkinson; pers. comm.). In the second, definitive, pass we used the gene set derived from this first MAKER iteration to train Augustus \cite{Stanke:2003eo} inside the MAKER pipeline for a second run, also including evidence from transcriptome to genome mapping obtained with GenomeThreader \cite{Gremme:2005ik}. Codon usage in \RCU, \TSP\ and \CEL\ was calculated using INCA (v2.1) \cite{Supek:2004et}. Results were then compared to data from \cite{Cutter:2006kz} (see Supplementary files 1 and 3).\\ 
We used Blast2GO  (Blast2GO4Pipe, v.2.5, January 2012 database issue) \cite{Conesa:2005hq} to annotate the gene set with Gene Ontology terms \cite{Ashburner:2000wc}, based on BLAST matches with expect values less than 1e$^-$$^5$ to the UniProt/SwissProt database (March 2012 snapshot), and domain annotations derived from the InterPro database \cite{Quevillon:2005jc}. Comparison of annotations between three nematode species (\RCU, \CEL\ and \TSP) and, as a reference outgroup, the holometabolous coleopteran arthropod \textit{Tribolium castaneum} was based on GO Slim data retrieved with Blast2GO. RNA genes were predicted using INFERNAL (v.1.0.2)\cite{Nawrocki:2009kr} and the Rfam database \cite{GriffithsJones:2004fq}, and tRNAscan-SE (v.1.3.1)\cite{Lowe:1997uc}.

\subsection*{Orthology Screen}
We inferred clusters of orthologous proteins between \RCU, \TSP\ and \CEL, and the beetle \textit{T. castaneum} using OrthoMCL (v.2.0.3) \cite{Li:2003en}. \TSP, \CEL \ and \TCA \ protein sets were downloaded from NCBI and WormBase (see Table S2) and redundancy screened with Cd-hit at the 99\% threshold. We selected an inflation parameter of 1.5 for MCL clustering (based on \cite{VanDongen:2000usb, vanDongen:2000us}) within OrthoMCL to generate an inclusive clusterings in our analysis likely to contain even highly diverged representatives from the four species. In analyses of selected developmental genes, clusters were manually validated using NCBI-BLAST+ \cite{Altschul:1990vt}. 
We affirmed the uniqueness of \CEL\ proteins identified as lacking homologues in the enoplean nematodes by comparing them to the \RCU\ proteome using BLAST. Those with no significant matches at all (all matches with E-values $>$ 1e$^-$$^5$) were classified as confirmed absent. Those having matches with E-values $<$ 1e$^-$$^5$ were investigated further by surveying the cluster memberships of the \RCU\ matches. If the \RCU\ protein was found to cluster with a different \CEL\ protein, the uniqueness to \CEL\ was again confirmed. If the \RCU\ protein did not cluster with an alternative \CEL\ protein, we reviewed the BLAST statistics (E-value, identity and sequence coverage) of the match and searched the GenBank non redundant protein database for additional evidence of possible orthology. Only if these tests yielded no indication of direct orthology was the \CEL\ protein designated absent from the enoplean set. Further details of the process are given in Supplementary file 5.\\
We identified the protein sequences of 1,725 genes differentially expressed in \CEL\ developmental stages \cite{Levin:2012vd} and selected, using our OrthoMCL clustering, those apparently lacking orthologues in \RCU\ and \TSP\ (verified as above). Using Wormbase (\url{http://www.wormbase.org}, release WS233) we surveyed the \CEL-restricted genes for their experimentally-defined roles in development.\\
Custom Perl scripts were used to group orthoMCL clusters on the basis of species membership patterns. The sets of clusters that contained (i) both \TSP\  and \RCU\ members but no \CEL\ members and (ii)  \TSP\  and \RCU\ and \TCA\ members but no \CEL\ members were surveyed for GO annotations enriched in comparison to the whole \CEL\ proteome (sets i and ii) and the \TCA\ proteome (set i), conducting Fisher's exact test as implemented in Blast2GO. To improve annotation reliability, these proteins were recompared (using BLAST) to the UniProt/SwissProt database and run through the Blast2GO pipeline in the same way as described above.  

\subsection*{Whole-mount in situ hybridization}
For in situ hybridisation we modified the freeze-crack procedure described previously for \CEL\ \cite{Seydoux:1995tk} and revised by Maduro et al. (2007; \url{http://www.faculty.ucr.edu/~mmaduro/resources.htm}). In particular to allow for reliable penetration of the durable \RCU\ egg envelopes we initially partly removed the protective layer by incubation in alkaline bleach solution (see \cite{Schulze:2008bc}). Digoxygenine-labeled sense and antisense RNA probes were generated from linearized pBs vectors (Stratagene, La Jolla, USA) containing a 400 bp fragment of \RCU\ \textit{mex-3} via run off \textit{in vitro} transcription with T7 or T3 RNA-polymerase according to the manufacturer's protocol (Roche, Mannheim, Germany). The concentration of the labeled probes was about 300 ng $\times$ ml$^-$$^1$.

\bigskip

\section*{Author's contributions}
PHS conceived study, assembled and annotated the genome, conducted analyses and wrote paper; MK conceived study, conducted analyses and wrote part of the paper; CK conceived part of the study, conducted analyses on developmental expression set and wrote part of the paper; GDK  helped with genome assembly and annotation; SK  helped with genome assembly and wrote/provided Perl scripts; JIRC analysed MEX-3 dataset; NAN analysed PAR dataset; DS analysed SEX determination dataset; KM conducted RNA sequencing and initial EST assembly; PH performed preparative laboratory experiments and conceived sequencing strategy; JA conceived sequencing strategy and conducted genome sequencing; PF helped with initial genome pre-assembly;  PN initiated study and conceived sequencing strategy; WKT conceived parts of study; MLB conceived study and wrote paper; ES initiated and conceived study and wrote paper

\section*{Funding}
This work was partly funded through the SFB 680: 
"Molecular Basis of Evolutionary Innovations".\\
Philipp H. Schiffer is funded by the \hbox{VolkswagenStiftung}
in the "F\"orderinitiative Evolutionsbiologie".\\
Gerogios D. Koutsovoulos is funded by a UK BBSRC Research Studentship and an Overseas Reasearch Studentship from the University of Edinburgh. 

\section*{Acknowledgements}
  \ifthenelse{\boolean{publ}}{\small}{
We are indebted to E. Platzer, Riverside, for the continuous supply with R. culcivorax nematodes. We thank J. Schulze, Cologne, for advice on nematode cultivation and C. Becker and K. Konrad for expert technical assistance in the genome sequencing experiments. We are also grateful to H. Oezden, Cologne for assistance with In-situ hybridisations. We thank J. Parkinson, Toronto, for providing a conserved NEMBASE4 protein set, Elizabeth Mart\'inez Salazar, Zacatecas, Mexico, for Feulgen C-value data and J. Burr, Vancouver, Canada for sharing preliminary results on phototaxis in \RCU.\\ 
Assemblies and other computations were conducted on the HPC cluster "CHEOPS" at the University of Cologne (\url{http://rrzk.uni-koeln.de/cheops.html}).}


\newpage
{\ifthenelse{\boolean{publ}}{\footnotesize}{\small}
 \bibliographystyle{RCU-genome-addon2111-b}  
  \bibliography{RCU-genome-addon2111-b}}     


\ifthenelse{\boolean{publ}}{\end{multicols}}{}


\newpage
  \subsection*{Figure 1}
     \begin{figure}[ht!]
\begin{center}
\includegraphics[width=6cm]{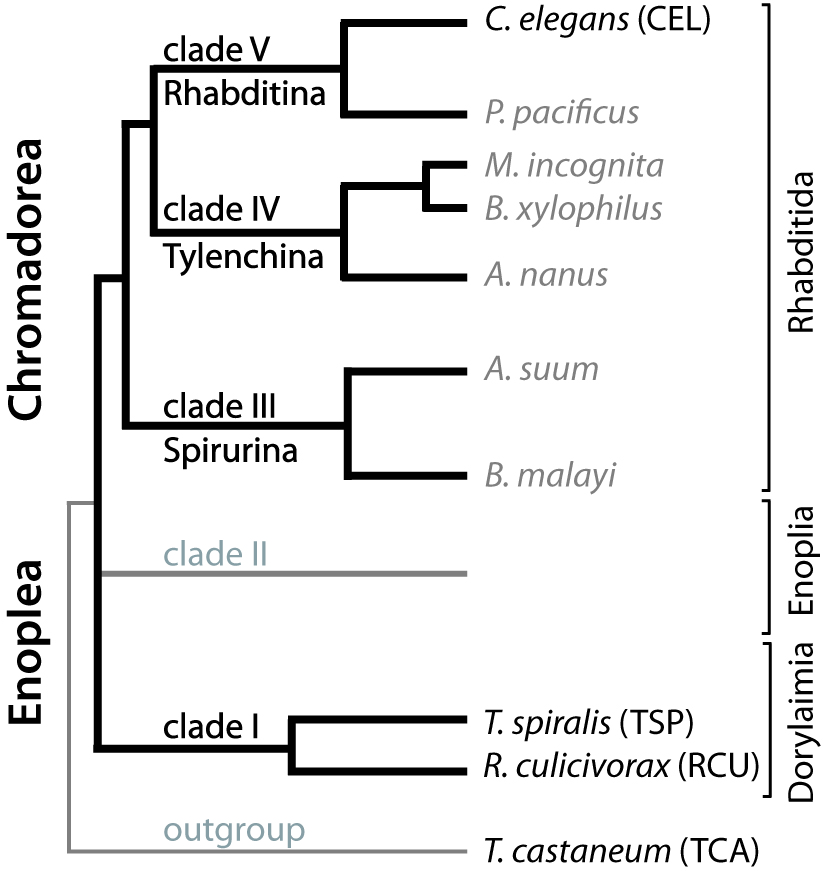}
\end{center}
\caption{
{\bf  A simplified phylogenetic tree of the phylum Nematoda.} The phylogeny, simplified from \cite{Blaxter:1998wj, Meldal:2007te}, emphasises the position of the main study species \RCU, \TSP\ and \CEL. The phylogenetic placements of species from Table 2 are given in grey. Currently no genomic data are available for Enoplia (Clade II). The order of branching of the basal nodes of Nematoda is unresolved.
\label{fig:1}}
\end{figure}

\newpage

\subsection*{Figure 2}
      \begin{figure}[ht!]
\begin{center}
\includegraphics[width=8cm]{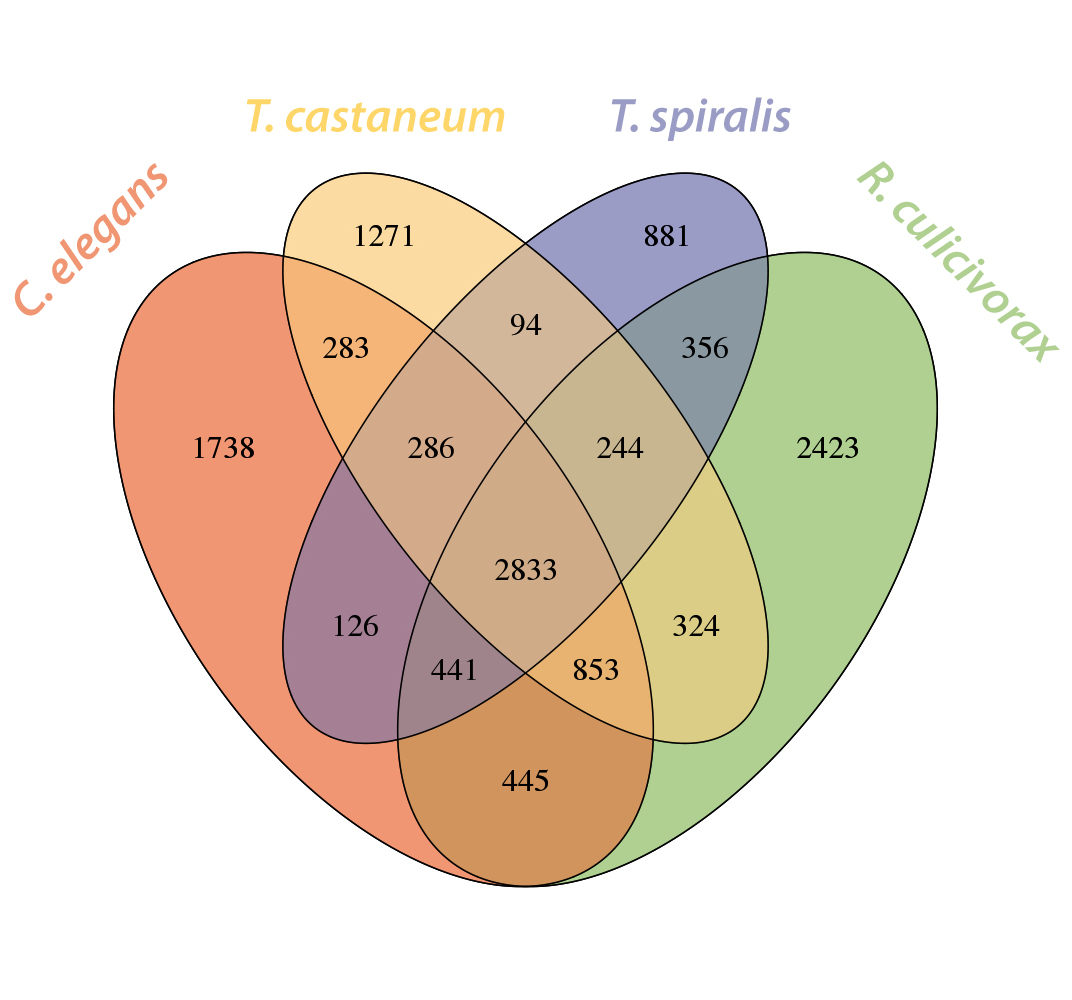}
\end{center}
\caption{
{\bf Clusters of homologous proteins.} Shared and species-unique clusters of homologous proteins from a comparison of the proteomes of \textit{Romanomermis culicivorax}, \textit{Trichinella spiralis}, \textit{Caenorhabditis elegans} and \textit{Tribolium castaneum} using OrthoMCL.} 

\end{figure}

\newpage

\newpage

  \subsection*{Figure 3}
     
      \begin{figure}[ht!]
\begin{center}
\includegraphics[width=15cm]{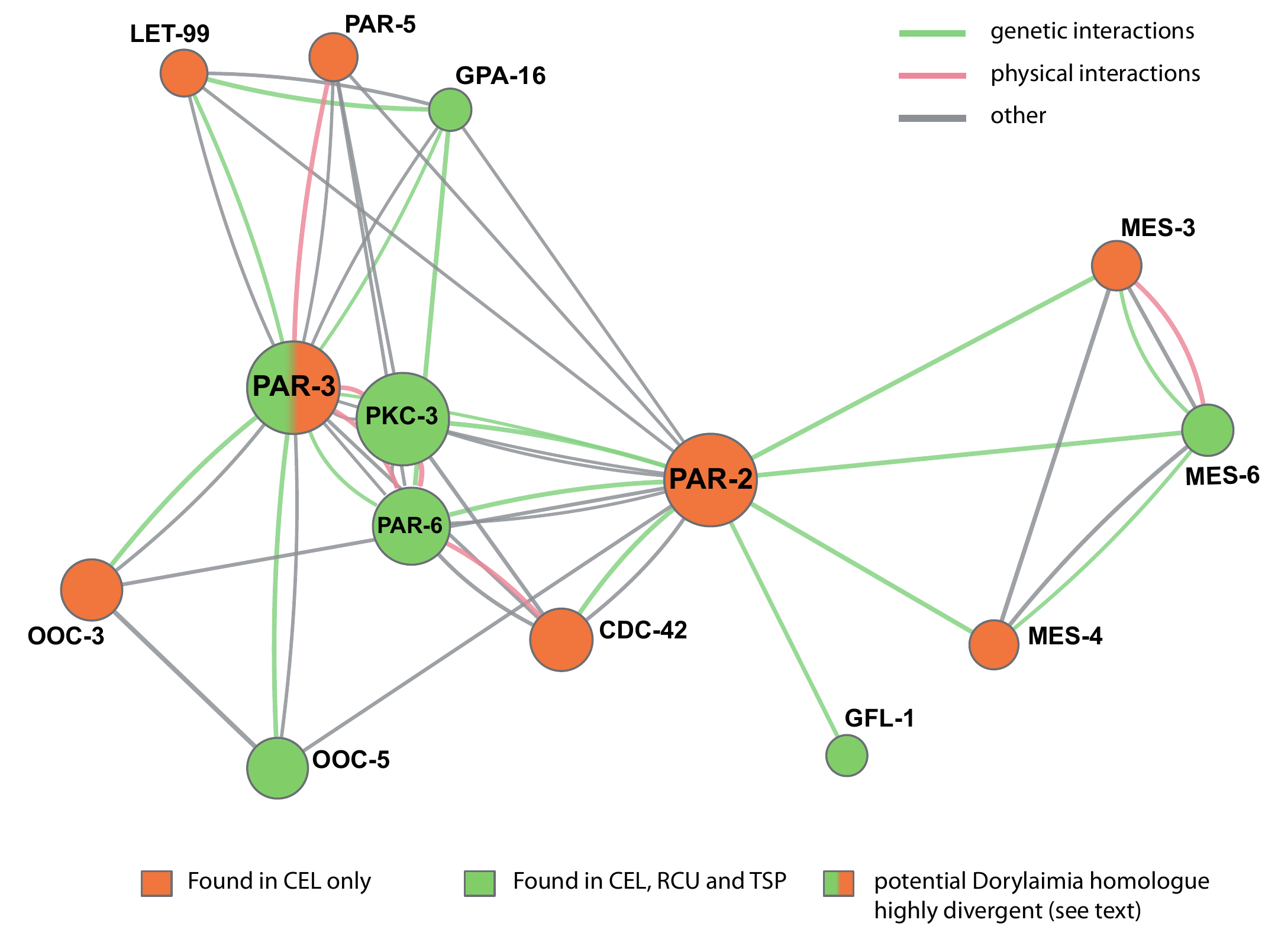}
\end{center}
\caption{\label{fig:4}
{\bf The network of proteins interacting with PAR-2 and PAR-3 in \textit{Caenorhabditis elegans} and their orthologues in \textit{Romanomermis culicivorax} and \textit{Trichinella spiralis}.} The network cartoon is based on the core polarity pathway extracted from WormBase, derived from both genetic and physical interactions. PAR-2 was missing from the dorylaim nematodes, as were the directly connected \textit{mes-3} and \textit{mes-4} genes. The {\RCU\ } PAR-3-like proteins was not retrieved as an orthologue of \CEL\ and \TSP\ PAR-3 proteins, but was identified employing sensitive sequence similarity search. See Table 3 for additional proteins interacting with PAR proteins and their presence-absence patterns.
}

\end{figure}

\newpage

  \subsection*{Figure 4}
      \begin{figure}[ht!]
\begin{center}
\includegraphics[width=12cm]{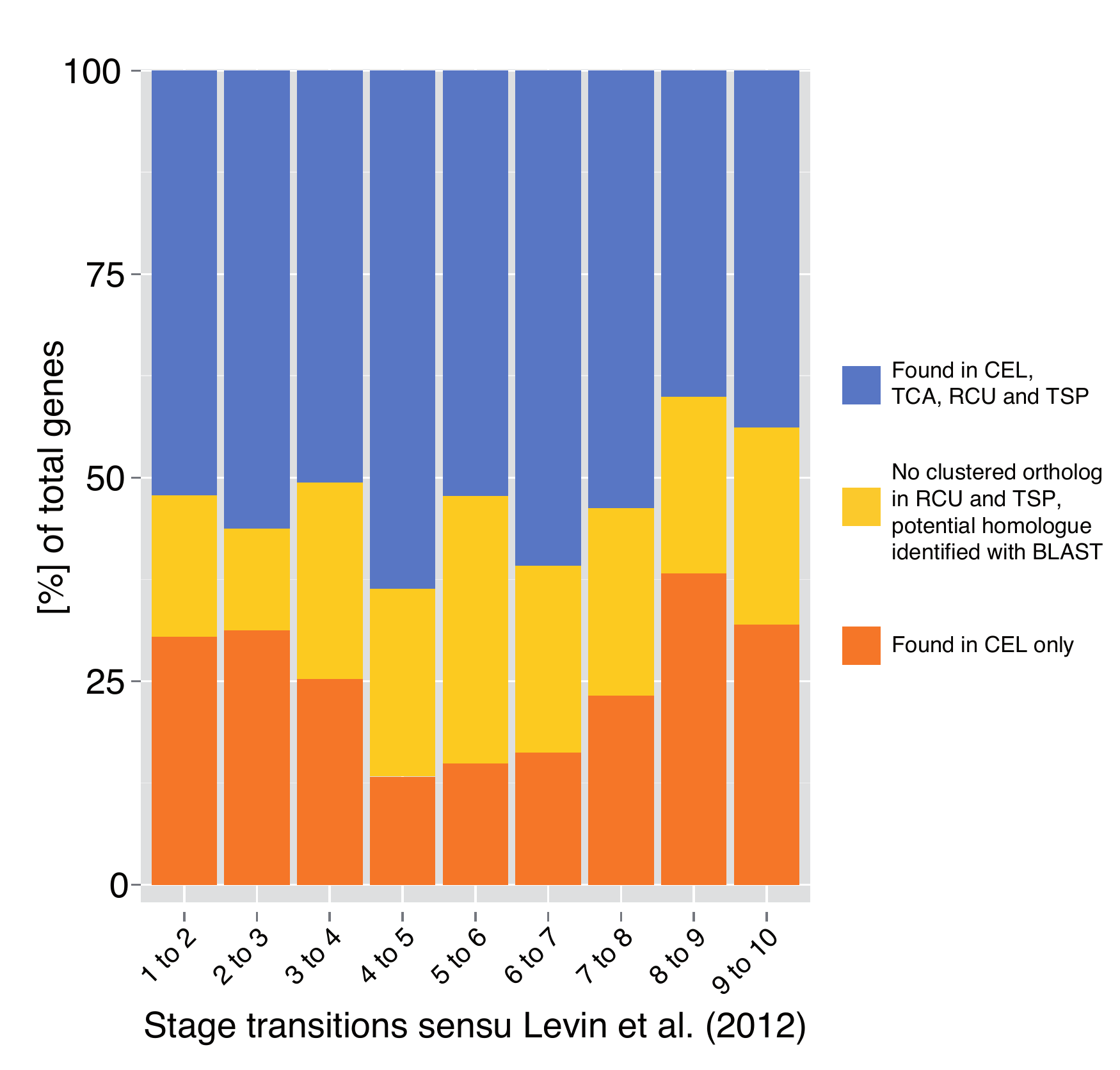}
\end{center}
\caption{
{\bf Many genes that are developmentally important in \textit{Caenorhabditis elegans} were not present in \textit{Romanomermis culicivorax} or \textit{Trichinella spiralis}.} \RCU\ and \TSP\ orthologues of the 1,725 genes identified as important in embryogenesis in an analysis of gene expression in \textit{Caenorhabditis} species \cite{Levin:2012vd} were sought. For each embryonic stage (1-10) in \CEL\ we calculated the proportion of these genes that were apparently unique to the genus \textit{Caenorhabditis}.}
\end{figure}

\newpage

  \subsection*{Figure 5}
     
      \begin{figure}[!h]
\begin{center}
\includegraphics[width=8cm]{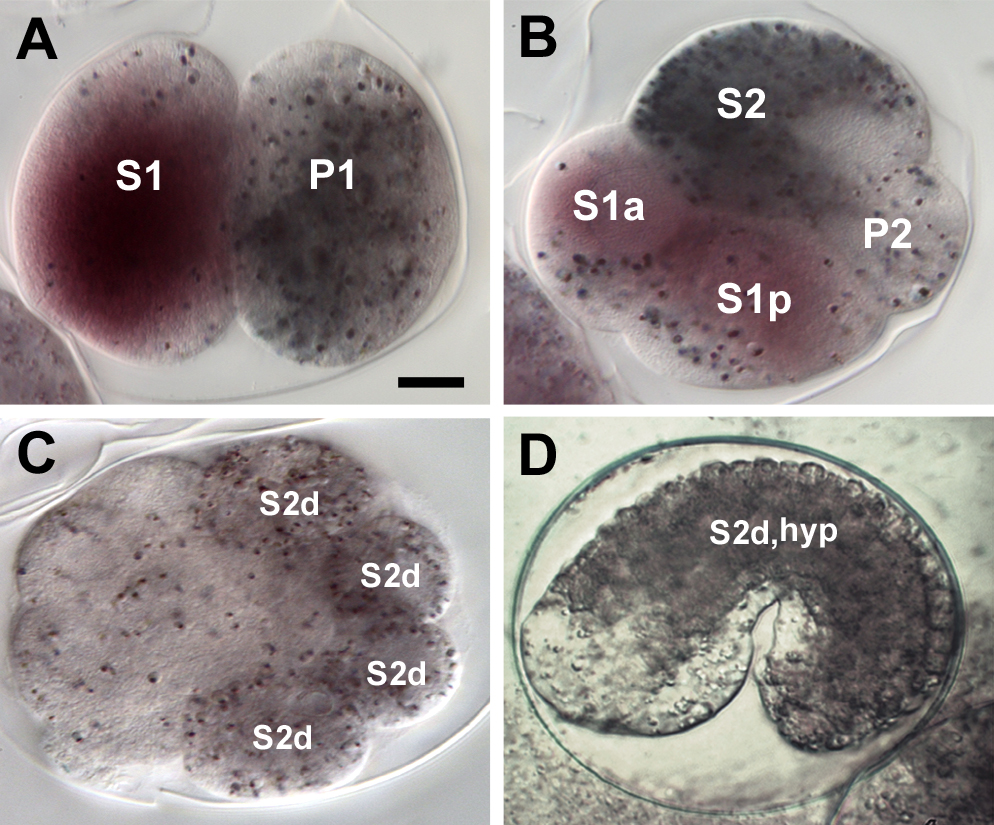}
\end{center}
\caption{
{\bf \textit{In situ} hybridisation revealing the pattern of distribution of \textit{mex-3} mRNA in early embryos of \textit{Romanomermis culicivorax}.} We used the \RCU\ \textit{mex-3} gene to prove application of the \textit{in situ} technique in this species and investigate the patterns of segregation of this maternal RNA in early development. The \RCU\ \textit{mex-3} expression pattern is similar to that of \CEL\ \cite{Draper:1996tq}. \RCU\ embryos contain dark pigment granules that are asymmetrically segregated in development. (A) At the 2-cell stage, maternal \textit{mex-3}  mRNA is detected in the S1 blastomere. The cytoplasmic pigment granules are predominantly in the P1 blastomere. (B) At the 4-cell stage, \textit{mex-3}  mRNA is detected in daughters of anterior S1 cell. Cytoplasmic pigment granules are predominantly in the S2 blastomere. (C) At a later stage (\textgreater 20 cells), \textit{mex-3} mRNA is absent. The pigment granules are found in descendants of S2 (S2d). (D) During early morphogenesis, the pigment granules are found in S2 descendants forming hypodermis, (S2d, hyp). (A-C) fixed embryos; (D) live embryo. Bar 10 $\mu$m. Orientation: anterior left.}

\end{figure}

\newpage


\section*{Tables}

\subsection*{Table 1-  Assembly and annotation statistics}
     \begin{table}[ht]
    \par
    \mbox{
      \begin{tabular}{>{\columncolor[gray]{0.9}}p{5cm}p{3cm}}
        \rowcolor[gray]{0.9} \bf{Metric} & \bf{Result} \\ \hline
        Contigs $>$100bp span & 267,342,457bp \\ \hline
        Scaffolds $>$500bp span & 322,765,761bp \\ \hline
        Num. contigs/scaffolds& 62,537 \\ \hline
        N50 contigs/scaffolds $>$500bp & 17,632 bp \\ \hline
        N50 scaffolds $>$500bp & 29,995bp \\ \hline
        Max contig length & 28,847bp  \\ \hline
         Max scaffold length&201,054bp \\ \hline
        Mean transcript length& 593bp \\ \hline
         Mean protein length & 190aa \\ \hline
        MAKER Augustus predictions & 12,026 proteins  \\ \hline
        MAKER SNAP predictions& 36,145 proteins \\ \hline
        Num. ESTs (isogroups) & 22,418 ESTs \\ \hline
        Mean EST length & 330bp \\ \hline
        80\% BLAT sequence coverage  & 21,204 ESTs   \\ \hline
        CEGMA compl. completeness & 75.40\% \\ \hline
        CEGMA Group 1 part. compl.& 81.82\% \\ \hline
        CEGMA Group 2 part. compl.& 91.07\% \\ \hline 
        CEGMA Group 3 part. compl. & 91.80\% \\ \hline
	CEGMA Group 4 part. compl. & 95.38\% \\ 	
        \end{tabular}
     }

    \end{table}

\subsection*{Table 2 - Genome statistics}
{\bf Repeat content of different nematode genomes appears not to be directly correlated with genome size. Re-calculation in selected genomes shows little deviance from published data (in parentheses){${^*}$} and thus indicates the validity of our inference for \RCU.}  
\\
{\small${^*}${For \textit{B. xylophilus} and \textit{M. incognita} only reference data is given as the same programs were used for initial inference (see references); \textit{A. suum} not re-calculated.}}\\

\begin{table}[ht]
    \par
\begin{tabular}{>{\columncolor[gray]{0.9}}p{2.5cm}p{2cm}p{2.5cm}p{2cm}p{2.2cm}p{1cm}p{1.5cm}}
\rowcolor[gray]{0.9}Species & Approximate${^\#}$\ \newline genome\ size & Estimated \ \newline Repeat content &Median${^\dag}$ \newline exon length& Median${^\dag}$ \newline intron length & GC \ \newline content& Source \\
\hline
\textit{C.elegans}&100Mb&17\% (16.5\%)& 145bp & 69bp&38\% &\cite{CelegansSequencingConsortium:1998wf, Stein:2003ks}\\
\hline
\PPA&165Mbp&15.3\% (17\%)&85bp&141bp& 42\% &\cite{Dieterich:2008bi, Kikuchi:2011fz}\\
\hline
\textit{A. suum} & 334Mb & 4.4\% & 144bp & 907bp &37.9\%& \cite{Jex:2011ew,Magrini:140vm}\\
\hline
\textit{B. malayi}&95Mb&16.5\% (15\%)&140bp&219bp&30\%&\cite{Ghedin:2007db}\\
\hline
\textit{B. xylophilus}&69Mb&22,5\%& 183bp&69bp&40\%&\cite{Kikuchi:2011fz}\\
\hline
\textit{M. incognita}&${\sim}$200Mb&36,7\%& 136bp&82bp & 31\% & \cite{Abad:2008kv} \\
\hline
\TSP&63Mb&19.8\% (18\%)&128bp&283bp& 34\% & \cite{Mitreva:2011ik}\\
\hline
\RCU&\textgreater 270Mb&48.2\%&161bp&405bp&36\%&this work \\
\end{tabular}
\\
{\small${^\#}${\textit{M. incognita} genome size given as 86Mbp in \cite{Abad:2008kv} has been re-estimated to about 200Mbp (E. Danchin pers. comm.).}}\\
{\small${^\dag}${Median lengths for \textit{A. suum} and \TSP\ were calculated in this work as these data are not given in the cited publications.}}\\
\end{table}

\newpage

 \newpage 
 \subsection*{Table 3 Presence and absence of selected \CEL\  proteins in Dorylaimia}
     \begin{table}[!h]
    \par
   \footnotesize{
      \begin{tabular}{@{} p{1.5cm}p{1.5cm}p{2.0cm} @{}}
\rowcolor[gray]{0.9} \bf{Protein}&  \bf{\TSP} & \bf{\RCU}\\ \hline
\multicolumn{2}{c}\mbox{\bf{Early asymmetry}} \\ \hline
\cellcolor[gray]{0.9}CDC-42&+&+\\
\cellcolor[gray]{0.9}PKC-3&+&+\\ 
 \cellcolor[gray]{0.9}GPR-1&+&+\\
 \cellcolor[gray]{0.9}GPR-2&+&+\\
\cellcolor[gray]{0.9} PAR-6&+&+\\
\cellcolor[gray]{0.9}MES-6&+&+\\ 
 \cellcolor[gray]{0.9}MES-3&-&-\\
 \cellcolor[gray]{0.9}MES-4&-&-\\ 
 \cellcolor[gray]{0.9}GFL-1&+&+\\
 \cellcolor[gray]{0.9}LET-70&+&+\\  \hline
\multicolumn{2}{c}\mbox{\bf{Axis formation}} \\ \hline
 \cellcolor[gray]{0.9}NUM-1&+&+\\
 \cellcolor[gray]{0.9}ZIM-1&-&-\\ 
 \cellcolor[gray]{0.9}MES-2&-&-\\
 \cellcolor[gray]{0.9}POS-1&-&-\\ 
 \cellcolor[gray]{0.9}SMA-6&+&+\\
 \cellcolor[gray]{0.9}SET-2&-&-\\ 
 \cellcolor[gray]{0.9}UBC-18&+&+\\
 \cellcolor[gray]{0.9}LET-99&-&-\\ 
 \cellcolor[gray]{0.9}OOC-3&-&-\\
 \cellcolor[gray]{0.9}OOC-5&+&+\\ 
 \cellcolor[gray]{0.9}GPA-16&+&+\\
 \cellcolor[gray]{0.9}PAR-5&-&-\\ 
\cellcolor[gray]{0.9}ATX-2&-&-\\
\cellcolor[gray]{0.9}MEX-5&-&-\\ 
\cellcolor[gray]{0.9} MEX-6&-&-\\
\cellcolor[gray]{0.9}UNC-120&-&-\\ 
 \cellcolor[gray]{0.9}NOS-2&-&-\\
 \cellcolor[gray]{0.9}OMA-1&-&-\\ 
 \cellcolor[gray]{0.9}RME-2&+&+\\
 \cellcolor[gray]{0.9}SPN-4&-&-\\  \hline
\multicolumn{2}{c}\mbox{\bf{Sex determination}} \\ \hline
\cellcolor[gray]{0.9}XOL-1&-&-\\
\cellcolor[gray]{0.9}HER-1&-&-\\ 
\cellcolor[gray]{0.9}SEX-1&+&+\\
\cellcolor[gray]{0.9}FOX-1&+&+\\ 
\cellcolor[gray]{0.9}SDC-1&-&-\\
\cellcolor[gray]{0.9}SDC-2&-&-\\ 
\cellcolor[gray]{0.9}SDC-3&-&-\\
\cellcolor[gray]{0.9}TRA-2&-&-\\ 
\cellcolor[gray]{0.9}FEM-1&+&+\\
\cellcolor[gray]{0.9}FEM-2&+&+\\  \hline
\multicolumn{3}{c}\mbox{\bf{Hypodermis and vulva formation}} \\ \hline
\cellcolor[gray]{0.9}AFF-1&-&-\\
\cellcolor[gray]{0.9}BAR-1&-&-\\ 
\cellcolor[gray]{0.9}CEH-2&-&-\\
\cellcolor[gray]{0.9}CEH-27&-&-\\ 
\cellcolor[gray]{0.9}GRL-15&-&-\\
\cellcolor[gray]{0.9}INX-5&-&-\\ 
\cellcolor[gray]{0.9}LIN-1&-&-\\
\cellcolor[gray]{0.9}PEB-1&-&-\\ 
\cellcolor[gray]{0.9}ELT-3&-&-\\
\cellcolor[gray]{0.9}ELT-1&+&+\\ 
\cellcolor[gray]{0.9}SMA-3&-&-\\
\cellcolor[gray]{0.9}SMA-5&-&-\\ 
\end{tabular}
}
\end{table} 
 \newpage

 \newpage 

\section*{Supplementary Files}

These will be available through the main author upon personal request in the preprint phase.

\subsection*{Supplementary file 1 --- Supplementary Figures and Tables}

\subsection*{Supplementary file 2 --- Fisher's exact test data}
    GO terms enriched in a set of protein clusters shared between Dorylaimia in comparison to (i) \CEL \ and (ii) \TCA\ proteomes.
  
\subsection*{Supplementary file 3 --- Codon usage in \RCU}
    Codon usage data.

\subsection*{Supplementary file 4 --- Levin data}
    Genes identified as being differentially expressed in \textit{Caenorhabditis} development by Levin et al. \cite{Levin:2012vd}.    

\subsection*{Supplementary file 5 --- Analysis of orthoMCL output by BLAST+}
	BLAST+ results for specific \CEL\ proteins not found in a cluster with Dorylaimia proteins.    

\end{bmcformat}
\end{document}